\documentclass[aps,twocolumn,showpacs,preprintnumbers,amsmath,amssymb,floatfix]{revtex4-2}

\usepackage{graphicx}
\usepackage{longtable}
\usepackage{dcolumn}
\usepackage{bm}
\usepackage{color}
\usepackage[normalem]{ulem}
\usepackage[colorlinks=true, urlcolor=blue, linkcolor=blue, citecolor=blue]{hyperref}
\usepackage[all]{hypcap}
\newcommand{\beq}{\begin{equation}}
\newcommand{\eeq}{\end{equation}}
\newcommand{\bea}{\begin{eqnarray}}
\newcommand{\eea}{\end{eqnarray}}

\begin{document}

\title{Correlation effects in two-dimensional $MX_2$ and $MA_2Z_4$ ($M$=Nb, Ta; $X$=S, Se, Te; $A$=Si, Ge; $Z$=N, P) cold metals: Implications for device applications}

\author{W. Beida$^{1,2}$} 
\author{E. \c{S}a\c{s}{\i}o\u{g}lu$^{3}$}\email{ersoy.sasioglu@physik.uni-halle.de}
\author{M. Tas$^{4}$}
\author{C. Friedrich$^{1}$}
\author{S. Bl\"{u}gel$^{1}$}
\author{I. Mertig$^{3}$}
\author{I. Galanakis$^{5}$}\email{galanakis@upatras.gr}

\affiliation{$^{1}$Peter Gr\"{u}nberg Institut, Forschungszentrum Julich and JARA, 52425 Julich, Germany \\
$^{2}$Physics Department, RWTH-Aachen University, 52062 Aachen, Germany \\
$^{3}$Institute of Physics, Martin Luther University Halle-Wittenberg, 06120 Halle (Saale), 
Germany \\
$^{4}$Department of Physics, Gebze Technical University, 41400 Kocaeli, Turkey \\
$^{5}$Department of Materials Science, School of Natural Sciences, University of Patras, 
GR-26504 Patra, Greece}

\date{\today}

\begin{abstract}

Cold metals, characterized by their distinctive band structures, hold promise for innovative electronic 
devices such as tunnel diodes with negative differential resistance (NDR) effect and field-effect 
transistors (FETs) with sub-60 mV/dec subthreshold swing (SS). In this study, we employ the $GW$ 
approximation and HSE06 hybrid functional to investigate the correlation effects on the electronic 
band structure of two-dimensional (2D) cold metallic materials, specifically focusing on $MX_2$ and 
$MA_2Z_4$ ($M$=Nb, Ta; $X$=S, Se, Te; $A$=Si, Ge; $Z$= N, P) compounds in 1H structure. These materials 
exhibit a unique band structure with an isolated metallic band around the Fermi energy, denoted as 
$W_{\mathrm{m}}$, as well as two energy gaps: the internal gap $E^{\mathrm{I}}_{\mathrm{g}}$ below the 
Fermi level and the external gap $E^{\mathrm{E}}_{\mathrm{g}}$ above the Fermi level. These three electronic 
structure parameters play a decisive role in determining the current-voltage ($I$-$V$) characteristics 
of tunnel diodes, the nature of the NDR effect, and the transfer characteristics and SS value of FETs. 
Our calculations reveal that both $GW$ and HSE06 methods yield consistent electronic structure properties 
for all studied compounds. We observed a consistent increase in both internal and external band gaps, 
as well as metallic bandwidths, across all pn-type cold metal systems. Notably, the internal band gap 
$E^{\mathrm{I}}_{\mathrm{g}}$ exhibits the most substantial enhancement, highlighting the sensitivity 
of these materials to correlation effects. In contrast, the changes in the metallic bandwidth $W_{\mathrm{m}}$ 
and external band gap $E^{\mathrm{E}}_{\mathrm{g}}$ are relatively modest. These findings offer valuable 
insights for designing and optimizing cold metal-based devices. Materials like NbSi$_2$N$_4$, NbGe$_2$N$_4$, 
and TaSi$_2$N$_4$ show particular promise for high-performance NDR tunnel diodes and sub-60 mV/dec SS FETs.

\end{abstract}

\maketitle

\section{Introduction}\label{sec1}

The experimental discovery of graphene through the exfoliation of graphite has sparked an unprecedented 
surge of interest in two-dimensional (2D) materials \cite{Castro2009}. The reduced dimensionality of 2D 
materials compared to bulk materials gives rise to unique quantum size effects, leading to novel physical 
and chemical properties \cite{Castro2009}. In particular, the confinement of electronic and optical properties 
within a 2D plane has opened up exciting possibilities for a wide range of technological applications. 
Among 2D materials, semiconductors such as transition metal dichalcogenides (TMDCs), including MoS$_2$ and 
WS$_2$, have garnered significant attention for their potential in next-generation electronic and optoelectronic 
devices \cite{Zhang2019}. Furthermore, 2D semiconductors are expected to play a crucial role in emerging 
technologies like valleytronics, spintronics, and energy harvesting, pushing the boundaries of modern 
semiconductor technology \cite{Kumbhakar2023,Zeng2018,Shanmugam2022}.

\begin{figure*}[!ht]
\begin{center}
\includegraphics[scale=0.19]{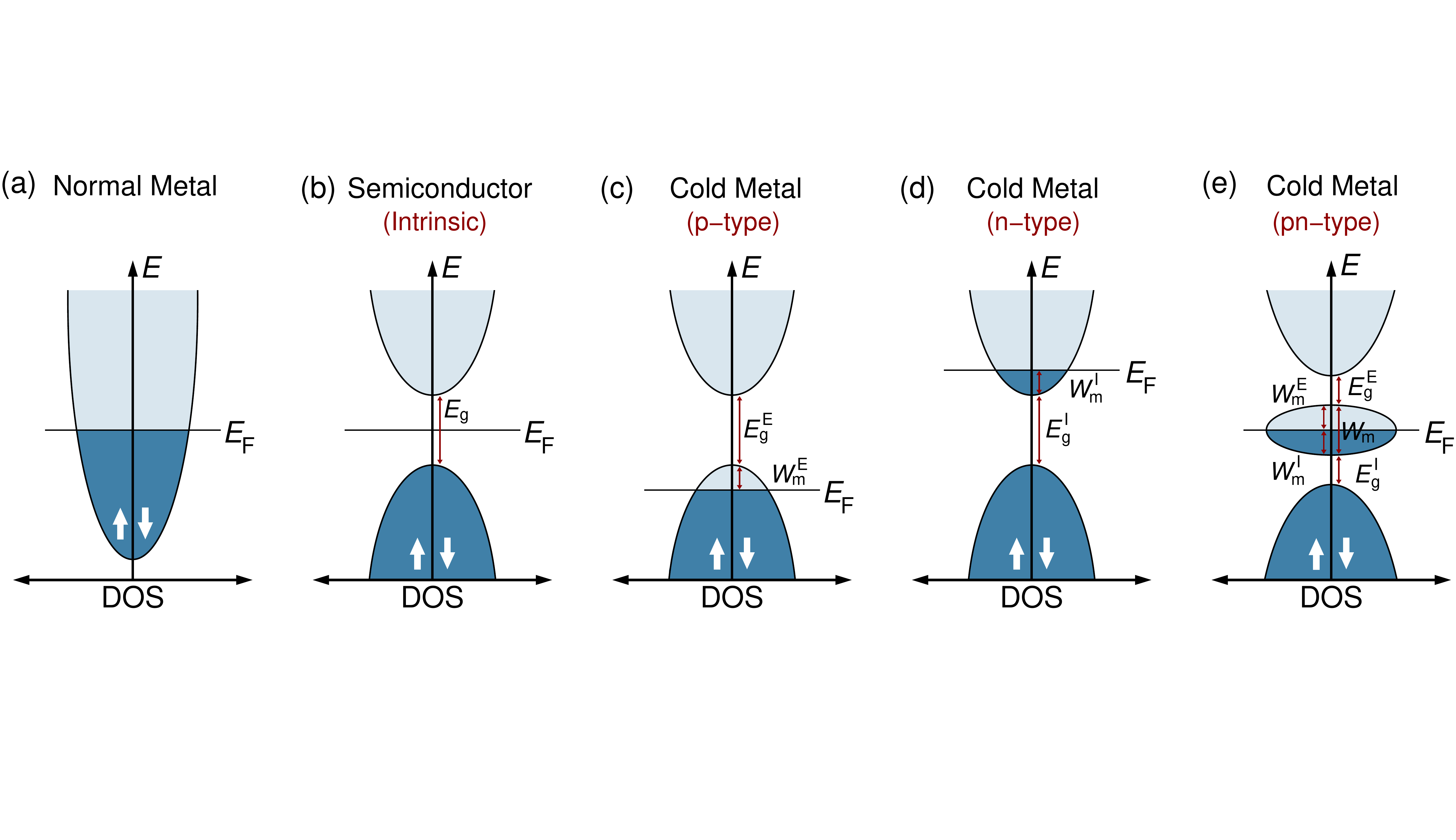}
\end{center}
\vspace*{-3.2cm} 
\caption{Schematic representation of the density of states (DOS) depicting: (a) a normal metal, (b) a 
semiconductor, (c) a p-type cold metal, (d) an n-type cold metal, and (e) a pn-type cold metal. The 
symbols $E_{\mathrm{g}}$, $E^{\mathrm{I}}_{\mathrm{g}}$, and $E^{\mathrm{E}}_{\mathrm{g}}$ corresponds to 
the band gap of the semiconductor, as well as the internal and external band gaps of cold metal, 
respectively. The width of the metallic band for the cold metal is designated by $W_{\mathrm{m}}$. 
The Fermi level is denoted by $E_{\mathrm{F}}$. For p- and n-type cold metals the distance between $
E_{\mathrm{F}}$ and the valence band maximum and conduction band minimum are denoted by $W_{\mathrm{m}}^{\mathrm{E}}$ 
and $W_{\mathrm{m}}^{\mathrm{I}}$, respectively.} 
\label{fig1}
\end{figure*}

Despite the promise of 2D semiconductors like TMDCs, the search for novel materials with exceptional 
electronic properties remains critical. Among these materials, ``cold metals'' have emerged as a class 
with distinctive electronic structures and functionalities \cite{Singh2021}. These materials are 
characterized by their unconventional band structures, which deviate significantly from traditional 
metals and semiconductors. As shown in Fig.\,\ref{fig1}, cold metals exhibit unique densities of states 
(DOS), featuring well-defined internal ($E^{\mathrm{I}}_{\mathrm{g}}$) and external ($E^{\mathrm{E}}_{\mathrm{g}}$) 
band gaps, along with a metallic band width ($W_{\mathrm{m}}$) that govern their electronic properties. 
Depending on the position of the Fermi level, cold metals can be classified into three types: p-type, 
n-type, and pn-type. In p-type and n-type cold metals, the Fermi level intersects the valence or conduction
bands, respectively, resulting in intrinsic conductivity without the need for doping. Of particular interest 
are pn-type cold metals, which feature an isolated metallic band near the Fermi level, flanked by energy
gaps, distinguishing them from conventional materials. In the literature, p- and n-type cold metals are 
also referred to as ``gapped metals'' \cite{Ricci2020,Khan2023,Malyi2020}.

Cold metals have already been investigated for a wide range of applications, from thermoelectrics 
and plasmonics to nanoelectronics and spintronics. In plasmonics, cold metals present an innovative 
solution to the issue of optical losses 
\cite{Zhang2024,khurgin2010search,gjerding2017band,da2020universal,song2021plasmons,gao2022ultra}. 
Traditional metals experience significant energy dissipation due to electronic transitions, limiting 
the performance of plasmonic devices. In contrast, cold metals, with their reduced density of states 
for scattering in the near-infrared region and the presence of energy gaps, provide an efficient 
alternative by suppressing optical losses. This low-energy dissipation enables the design of high-performance 
plasmonic devices \cite{Yu2019}. Moreover, cold metals show great potential in thermoelectric applications
due to their unique electronic structure \cite{Ricci2020}. Unlike conventional metals, which exhibit 
a relatively constant density of states near the Fermi level, cold metals display a sharp decrease in the 
density of states around the partial energy gap. This leads to an enhanced Seebeck coefficient, a key 
factor in efficient thermoelectric conversion. Combined with their high electrical conductivity, cold 
metals offer an optimized balance, positioning them as strong candidates for advanced thermoelectric 
materials.

While cold metals hold significant promise in thermoelectric and plasmonic applications, their 
potential in nanoelectronics is equally compelling. These materials enable innovative device 
concepts such as steep-slope field-effect transistors (FETs) and negative differential resistance 
(NDR) tunnel diodes with ultra-high peak-to-valley current ratios (PVCR) \cite{Sasioglu2023,Yin2022,Sasioglu2024}. 
By harnessing the unique electronic properties of cold metals, these devices could outperform 
conventional semiconductor technologies,  potentially leading to more energy-efficient electronics.

\section{Motivation and Aim}\label{sec2}

One of the key advantages of cold metals in nanoelectronics lies in their lack of band tails, a common 
issue in traditional p- and n-type semiconductors caused by doping and fluctuations in doping concentrations.
Band tails have been extensively studied and are known to degrade the performance of devices such as tunnel 
FETs and Esaki tunnel diodes \cite{efros1974density,chakraborty2001density,kane1985band,van1992theory,sant2017effect,bizindavyi2018band,schenk2020tunneling}.
In tunnel FETs, band tails increase the subthreshold slope (SS), while in Esaki diodes, they reduce the 
peak-to-valley current ratio (PVCR). Cold metals, by contrast, are free of band tails, providing an inherent 
advantage in the design of advanced nanoelectronic devices, such as steep-slope transistors and NDR tunnel 
diodes \cite{Wang2022}.

Cold metals are theoretically proposed to overcome the thermionic limit of 60 mV/decade SS in traditional FETs, which 
limits the reduction of power consumption while maintaining switching speed \cite{Liu2020}. The primary challenge in 
conventional transistors is the presence of high-energy hot electrons that contribute to this limit. In experimental 
studies, graphene has been demonstrated as a potential solution due to its ability to suppress high-energy electrons, 
thanks to its unique band structure \cite{qiu2018dirac,tang2021steep}. However, recent theoretical works suggest that 
cold metals offer an even better alternative than graphene \cite{Liu2020,Zhang2024}. Their intrinsic band gaps above 
the Fermi level act as energy filters for high-energy hot electrons, selectively allowing only low-energy cold electrons 
to participate in transport. This energy filtering mechanism could enable sub-60 mV/dec switching, significantly improving 
energy efficiency by reducing leakage currents and sharpening the switching curve. As a result, cold-metal-based FETs 
are projected to achieve faster switching at lower operating voltages, making them suitable for future energy-efficient 
computing systems.

\begin{figure*}
\begin{center}
\includegraphics[scale=0.22]{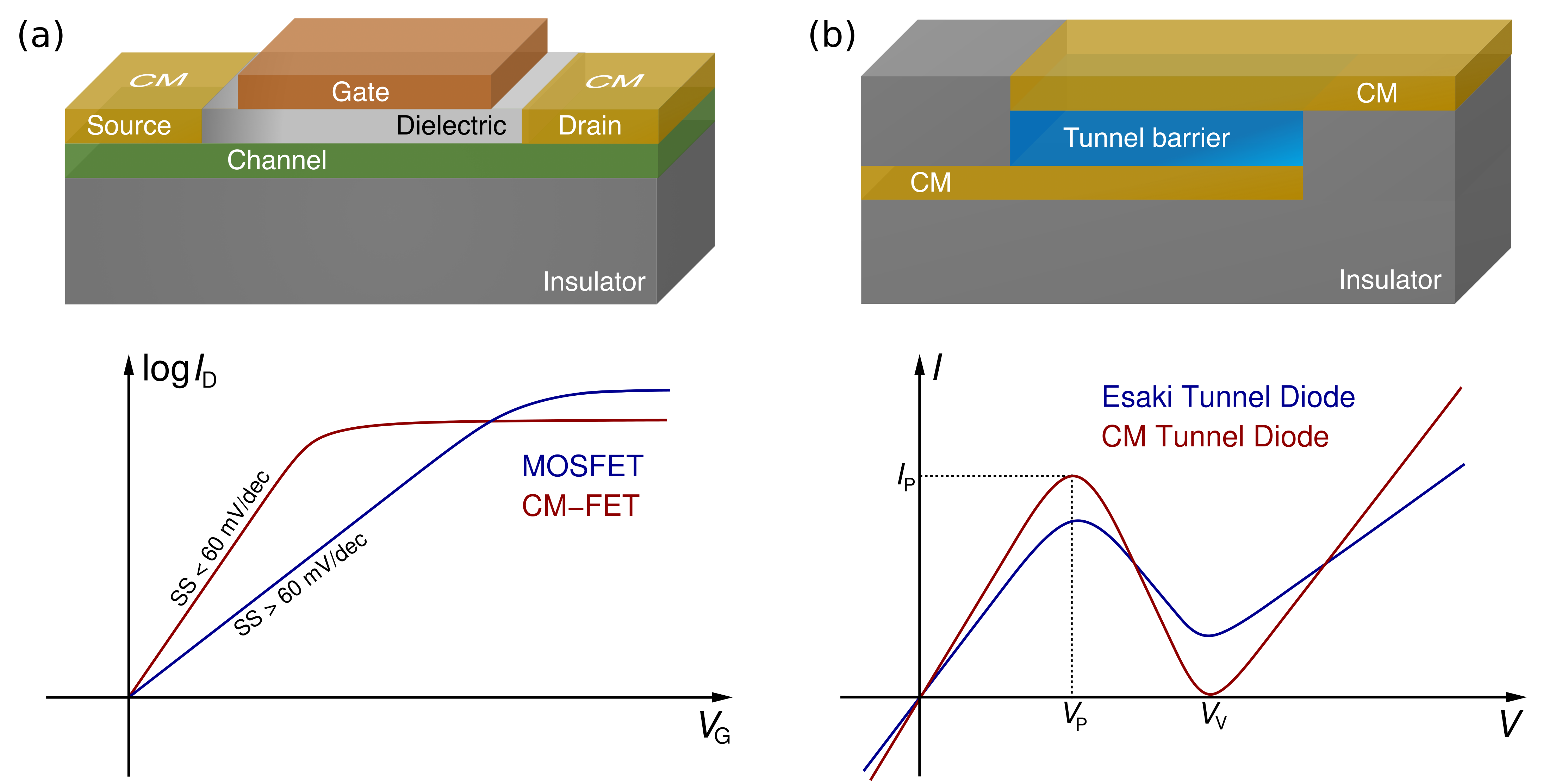}
\end{center}
\vspace*{-0.4cm} 
\caption{(a) Schematic illustration of a two-dimensional vertical field-effect transistor (FET) utilizing cold metal 
(CM) electrodes, along with its transfer characteristics, including a comparison with a conventional MOSFET. The 
subthreshold slope (SS) is indicated. (b) Schematic illustration of a two-dimensional vertical negative differential 
resistance (NDR) tunnel diode based on cold metals, accompanied by its current-voltage ($I$-$V$) characteristics, 
with a comparison to the $I$-$V$ curves of an Esaki tunnel diode.}
\label{fig2}
\end{figure*}

In Fig. 2(a), we present the schematic transfer characteristics of an FET utilizing cold metal source and drain electrodes, 
comparing it with a conventional MOSFET. The cold metal FET demonstrates a steeper subthreshold slope due to the energy 
filtering effect from the band gaps in cold metals. This comparison illustrates the potential of cold metal FETs to achieve 
better energy efficiency and sharper switching behavior than traditional MOSFETs, which are constrained by the thermionic 
limit. The theoretical promise of cold metals in FETs is supported by ab-initio simulations based on density functional 
theory (DFT). Simulations of cold-metal FETs using 2D materials like NbSe$_2$ and NbTe$_2$ have demonstrated subthreshold 
slopes below 60 mV/decade at room temperature \cite{Liu2020,Yin2022,Wang2022}. Additionally, these simulations show an 
NDR effect with ultra-high PVCR values, further emphasizing the potential of cold metals in electronic device applications \cite{Sasioglu2023,Yin2022,Wang2022}. 
While graphene-based transistors have already been experimentally demonstrated, cold metals offer a pathway to even more 
efficient steep-slope transistors. In addition to FETs, cold metals show potential in NDR tunnel diodes. By selecting 
appropriate pn-type cold metal electrodes, the tunnel diode can exhibit either an N-type (NbS$_2$/h-BN/NbS$_2$) or $\Lambda$-type 
(AlI$_2$/MgI$_2$/AlI$_2$ NDR effect \cite{Sasioglu2023}. In Fig.\,\ref{fig2}(b), we present schematic $I$-$V$ characteristics 
of a cold metal tunnel diode, highlighting the sharpness of the N-type NDR effect compared to a conventional Esaki diode.

Given the promising potential of cold metals in nanoelectronic devices, this study aims to investigate the effect of
electronic correlations on their electronic band structures. By employing the state-of-the-art $GW$ approximation and 
HSE06 hybrid functional, we explore changes in the internal ($E^{\mathrm{I}}_{\mathrm{g}}$)  and external ($E^{\mathrm{E}}_{\mathrm{g}}$)
band gaps, as well as the metallic bandwidth ($W_{\mathrm{m}}$), in $MX_2$ and $MA_2Z_4$ ($M$=Nb, Ta; $X$=S, Se, Te; $A$=Si, 
Ge; $Z$=N, P) compounds in the 1H structure. These electronic structure parameters are crucial for understanding the performance 
of cold metal-based tunnel diodes and FETs. Our study focuses on how these parameters vary across different material compositions 
and the influence of correlation effects. By understanding how these correlation effects impact the electronic properties, we aim 
to offer insights into the design and optimization of cold metal-based devices, such as steep-slope transistors and NDR 
tunnel diodes, contributing to the development of next-generation, energy-efficient nanoelectronics.

\section{Computational Methods}

\begin{figure*}
\begin{center}
\includegraphics[scale=0.275]{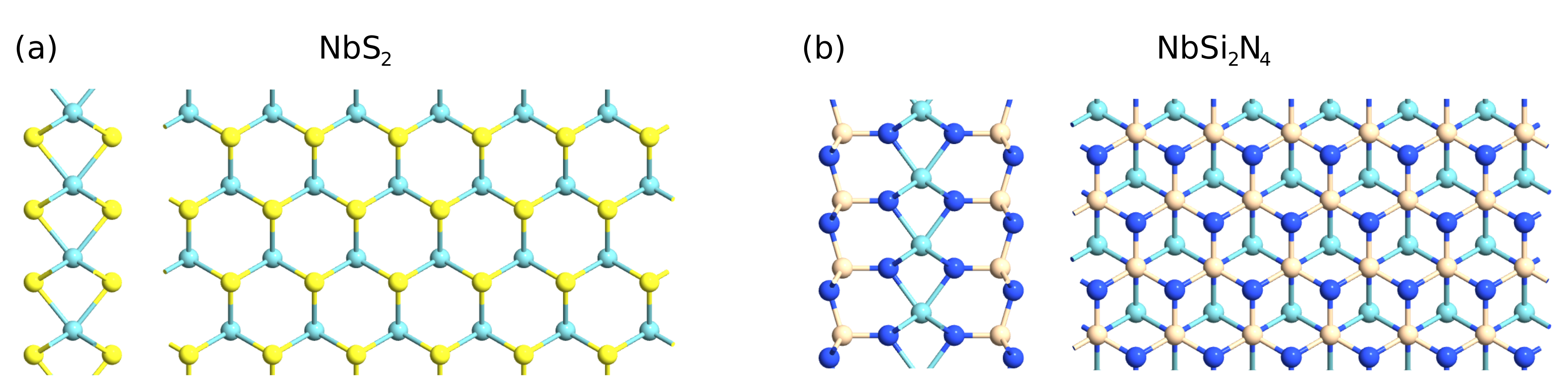}
\end{center}
\vspace*{-0.3cm} 
\caption{Side and top views of the two-dimensional crystal structure of monolayer 1H (a) NbS$_2$ and (b) NbSi$_2$N$_4$.
Both materials exhibit a hexagonal lattice, with each layer consisting of a transition metal (Nb) sandwiched between 
two chalcogen (S) atoms in NbS$_2$, or between pnictogen atoms (Si, N) in NbSi$_2$N$_4$. The 1H phase is characterized 
by its trigonal prismatic coordination of the transition metal atom.} 
\label{fig3}
\end{figure*}

\subsection{Crystal Structure}

The $MX_2$ and $MA_2Z_4$ compounds studied in this work crystallize in the 1H structure, a common phase 
for TMDCs. The 1H structure consists of layers of transition metal atoms (M) sandwiched between layers of 
chalcogenide (X) or pnictogen (A, Z) atoms. Fig.\,\ref{fig3} illustrates the crystal structure of these 
compounds. $MX_2$ compounds have been experimentally synthesized in both 1H and 1T structures \cite{Xiao2019,Xiao2023}. 
However, for this study, we focus on the 1H structure because it exhibits cold metallic behavior and is the most 
stable phase, while the 1T structure displays normal metallic characteristics. The $MA_2Z_4$ compounds, proposed 
in Ref.\,\onlinecite{Wang2021}, are a new family of van der Waals layered materials. These compounds consist of 
a $MZ_2$ layer intercalated between two $AZ$ monolayers, as shown in Fig.\,\ref{fig3} (b). Some members of this 
family, such as MoSi$_2$N$_4$, were experimentally synthesized before theoretical studies \cite{Hong2020}. This 
initial  computational study led to the prediction of 72 
thermodynamically and dynamically stable $MA_2Z_4$ compounds \cite{Wang2021}. For our calculations, 
we included six  compounds from 
Ref. \onlinecite{Wang2021} which are cold metals: 
(Nb,Ta)(Si,Ge)$_2$N$_4$ and (Nb,Ta)Si$_2$P$_4$. The lattice constants are 
taken from the  Computational 2D Materials Database (C2DB)  for the $MX_2$ compounds \cite{C2DB,Rasmussen2015,Haastrup2018} 
while for the $MA_2Z_4$ compounds the lattice constants were taken from Ref.\,\onlinecite{Wang2021} and are presented 
in Table\,\ref{table1}.

\subsection{Computational Details}\label{sec3b}

To investigate the electronic band structures of the $MX_2$ and $MA_2Z_4$ compounds, we employed two 
first-principles electronic structure methods. The first approach utilized the \textsc{QuantumATK} 
software package ~\cite{QuantumATK,QuantumATKb}, which implements both linear combinations of atomic 
orbitals (LCAO) and Plane Waves (PW) as the basis sets along with norm-conserving PseudoDojo pseudopotentials ~\cite{VanSetten2018}. 
We initially employed the Perdew-Burke-Ernzerhof (PBE) parameterization of the generalized-gradient 
approximation (GGA) for exchange-correlation energy ~\cite{Perdew1996}, a standard functional in 
materials simulations. For improved accuracy in describing exchange interactions, we also applied 
the Heyd–Scuseria–Ernzerhof (HSE06) hybrid functional ~\cite{HSE03,HSE06}, which mixes exact 
short-range Hartree-Fock exchange with long-range GGA exchange. While hybrid functionals like HSE06 are particularly 
suited for improving band gap predictions, their predefined mixing coefficients may not be universally 
optimal across all material systems. Nevertheless, given the long-range nature of Coulomb interactions 
in 2D cold metals \cite{Ramezani2024}, HSE06 provides an appropriate level of accuracy for capturing the electronic 
properties of these materials. In all cases, the correlation energy was treated using the PW92 correlation 
functional ~\cite{PW92}. To simulate the 2D monolayer limit, a vacuum region of 25\,\AA \,  was added in the 
1H structure unit cell to prevent inter-layer interactions, and a $18 \times 18 \times 1$ Monkhorst-Pack 
k-point grid \cite{Monkhorst1976} was employed in all self-consistent calculations.

To capture correlation effects more accurately, we performed many-body perturbation theory calculations 
within the $GW$ approximation. Ground-state properties were first computed using the full-potential 
linearized augmented-plane-wave (FLAPW) method implemented in the FLEUR code ~\cite{FLEUR}, using PBE 
for exchange and PW92 for correlation. Angular momentum and plane-wave cutoff parameters were set to 
$l_{\textrm{max}}=8$ inside muffin-tin spheres and $k_{\textrm{max}}=4.5$ bohr$^{-1}$ in the interstitial 
region. The Brillouin zone was sampled using an $18 \times 18 \times 1$ \textbf{k}-point grid.

Subsequently, one-shot $GW$ calculations were performed using the SPEX code ~\cite{SPEX,SPEX2}. In this approach, 
the off-diagonal elements in the self-energy operator $\Sigma_\sigma(E_{n\bf{k}\sigma})$ were neglected, 
and the expectation values of the local exchange-correlation potential $V_\sigma^{\textrm{XC}}$ were 
subtracted to avoid double-counting. The Kohn-Sham (KS) single-particle wavefunctions $\varphi_{n\bf{k}\sigma}^{\textrm{KS}}$ 
were treated as approximations to quasiparticle (QP) wavefunctions. Hence, the QP energies $E_{n\bf{k}\sigma}$ 
were computed as a first-order perturbation correction to the KS values $E_{n\bf{k}\sigma}^{\textrm{KS}}$ as
$ E_{n\bf{k}\sigma}=E_{n\bf{k}\sigma}^{\textrm{KS}}+\left\langle \varphi_{n\bf{k}\sigma}^{\textrm{KS}}| \Sigma_\sigma(E_{n\bf{k}\sigma}) - V_\sigma^{\textrm{XC}}|\varphi_{n\bf{k}\sigma}^{\textrm{KS}}\right\rangle$, 
where $n$, \textbf{k}, and $\sigma$ represent the band index, Bloch vector, and electron spin, respectively. 
The dynamically screened Coulomb interaction $W$ was expanded using a mixed product basis set, with contributions 
from both the local atom-centered muffin-tin spheres and plane waves in the interstitial region ~\cite{Kotani}. 
The cutoff parameters for the mixed product basis were set to $L_{\textrm{max}}=4$ and $G_{\textrm{max}}=4$ bohr$^{-1}$. 
A consistent computational cell and k-point grid ($18 \times 18 \times 1$) were used across all codes. Relativistic 
corrections were treated at the scalar-relativistic level for the valence states, while the core states were calculated 
using the full Dirac equation.

\begin{table*}
\caption{\label{table1} Lattice constants, cold metal type according to the $GW$-based calculations,  internal energy gap $E^{\mathrm{I}}_{\mathrm{g}}$, the external energy 
gap $E^{\mathrm{E}}_{\mathrm{g}}$, and the metallic bandwidth $W_{\mathrm{m}}$ (see Fig. \ref{fig1} for the definitions) using both 
FLEUR and \textsc{QuantumATK} codes in conjunction with the PBE and HSE06 functionals as well as the $GW$ approximation. For 
the \textsc{QuantumATK} code we present both the values obtained using 
the LCAO and the PW basis sets (the latter in parenthesis).
Lattice constants for
$MX_2$ compounds are  taken from the  Computational 2D 
Materials Database (C2DB)  \cite{C2DB,Rasmussen2015,Haastrup2018}, while those 
for $MA_2Z_4$ compounds are from Ref. \onlinecite{Wang2021}.}
\begin{ruledtabular}
\begin{tabular}{lclcccccc| cccccc} 
&  &   \multicolumn{6}{c}{ FLEUR/SPEX} &   \multicolumn{6}{c}{\textsc{QuantumATK}} \\
\cline{4-9} \cline{10-15} 

Comp. & $a_0$ & CM & \multicolumn{2}{c}{$E^{\mathrm{I}}_{\mathrm{g}}$(eV)} & \multicolumn{2}{c}{$E^{\mathrm{E}}_{\mathrm{g}}$(eV)} & \multicolumn{2}{c}{$W_{\mathrm{m}}$(eV)} 
& \multicolumn{2}{c}{$E^{\mathrm{I}}_{\mathrm{g}}$(eV)} & \multicolumn{2}{c}{$E^{\mathrm{E}}_{\mathrm{g}}$(eV)} & \multicolumn{2}{c}{$W_{\mathrm{m}}$(eV)}  \\ 
\cline{4-5}  \cline{6-7} \cline{8-9} \cline{10-11} \cline{12-13} \cline{14-15} 
 &  (\AA) & Type & $GW$ & PBE   & $GW$ & PBE & $GW$ & PBE &  HSE06 & PBE & HSE06 & PBE & HSE06 & PBE \\
\hline
NbS$_2$        &  3.35  & pn &  0.95 & 0.57  & 1.42 & 1.21  & 1.50 & 1.19       & 0.84 (0.86) & 0.45 (0.46) & 1.27 (1.29) & 1.15 (1.17) & 1.61 (1.60) & 1.24 (1.21)  \\ 
NbSe$_2$       &  3.47  & pn &  0.72 & 0.34  & 1.59 & 1.42  & 1.24 & 0.90       & 0.69 (0.63) & 0.30 (0.30) & 1.40 (1.49) & 1.28 (1.33) & 1.33 (1.31) & 0.98 (0.95)  \\
NbTe$_2$       &  3.70  & p  &  0.00 & 0.00  & 1.21 & 1.13  & 1.55 & 0.91       & 0.21 (0.14) & 0.00 (0.00) & 1.42 (1.38) & 1.16 (1.14) & 1.13 (1.22) & 0.92 (0.94)  \\
TaS$_2$        &  3.34  & pn &  1.22 & 0.67  & 1.59 & 1.35  & 1.78 & 1.40       & 0.98 (0.99) & 0.57 (0.58) & 1.39 (1.42) & 1.28 (1.31) & 1.89 (1.83) & 1.45 (1.41)  \\
TaSe$_2$       &  3.47  & pn &  0.49 & 0.38  & 1.70 & 1.47  & 1.47 & 1.11       & 0.84 (0.72) & 0.43 (0.39) & 1.49 (1.53) & 1.38 (1.40) & 1.55 (1.52)  & 1.14 (1.14) \\
TaTe$_2$       &  3.71  & pn &  0.16 & 0.08  & 1.22 & 1.03  & 1.43 & 1.11       & 0.29 (0.17) & 0.00 (0.00) & 1.28 (1.26) & 1.02 (1.02) & 1.32 (1.40) & 1.11 (1.13)  \\
NbSi$_2$N$_4$  &  2.97  & pn &  1.61 & 1.16  & 1.76 & 1.76  & 1.61 & 1.21       & 1.74 (1.74) & 1.11 (1.12) & 2.07 (2.09) & 1.73 (1.77) & 1.66 (1.65) & 1.27 (1.26)  \\
NbGe$_2$N$_4$  &  3.09  & pn &  1.63 & 1.04  & 1.20 & 1.00  & 1.29 & 1.09       & 1.60 (1.60) & 1.04 (1.04) & 1.31 (1.31) & 1.05 (1.04) & 1.41 (1.41) & 1.10 (1.10)  \\
NbSi$_2$P$_4$  &  3.53  & p  &  0.00 & 0.00  & 0.38 & 0.61  & 1.76 & 1.22       & 0.10 (0.11) & 0.00 (0.00) & 0.78 (0.80) & 0.55 (0.61) & 1.59 (1.58) & 1.23 (1.22)  \\
TaSi$_2$N$_4$  &  2.97  & pn &  1.70 & 1.25  & 1.47 & 1.52  & 1.83 & 1.46       & 1.93 (1.93) & 1.29 (1.29) & 1.66 (1.68) & 1.42 (1.46) & 1.97 (1.96) & 1.47 (1.46)  \\
TaGe$_2$N$_4$  &  3.08  & pn &  1.69 & 1.25  & 1.02 & 0.75  & 1.54 & 1.17       & 1.79 (1.78) & 1.28 (1.26) & 1.02 (1.01) & 0.82 (0.81) & 1.59 (1.60) & 1.18 (1.19)  \\
TaSi$_2$P$_4$  &  3.53  & p  &  0.00 & 0.04  & 0.18 & 0.35  & 1.94 & 1.48       & 0.30 (0.32) & 0.04 (0.06) & 0.49 (0.51) & 0.30 (0.36) & 1.83 (1.82) & 1.49 (1.48)  \\
\end{tabular}
\end{ruledtabular}
\end{table*}

\section{Results and discussion}

We study two families of 2D compounds in the 1H structure: transition metal dichalcogenides (TMDCs) (Nb, Ta)(S, Se, Te)$_2$ 
and layered compounds (Nb, Ta)(Si, Ge)$_2$(N, P)$_4$. Table\,\ref{table1} presents the cold metal character for each compound, 
alongside the widths of the internal and external energy gaps and the metallic bandwidth, as calculated using the PBE and HSE06 
functionals, as well as the $GW$ approximation. 

The choice of the HSE06 functional and $GW$ approximation is supported by prior studies on similar cold metals and TMDCs. For 
instance, SrVO$_3$ is a well-studied n-type cold metal where $GW$ and dynamical mean-field theory methods significantly reduce 
the metallic bandwidth, as demonstrated in Ref. \onlinecite{Sakuma2013}. However, the behavior of energy gaps within the $GW$ approximation 
compared to standard GGA calculations varies across studies. Regarding the HSE06 functional, a recent work \cite{Ramezani2024} 
reported the effective Coulomb interaction parameters for a broad set of $MX_2$ ($M$ = Mo, W, Nb, Ta; $X$ = S, Se, Te) compounds 
across the 1H, 1T, and 1T' phases, finding unconventional Coulomb screening in the 1H structure. This unconventional screening 
suggests that 2D cold metals do not screen Coulomb interactions as effectively as normal metals, thereby justifying the use of the 
HSE06 functional for these systems. While $GW$ is theoretically more accurate, HSE06 still provides reliable insights into the 
electronic properties of 2D cold metals.

\subsection{PBE functional}

Before presenting our HSE06 and $GW$ results, we first validated the consistency of our 
findings, given that two different codes based on different electronic band structure  methods (FLAPW-based FLEUR code  and LCAO-based or PW-based \textsc{QuantumATK} code) were used, as previously 
discussed. To ensure reliability, both methods should produce closely matching results for 
the studied compounds when employing the PBE functional. Table\,\ref{table1} lists the 
computed internal and external band gaps ($E^{\mathrm{I}}_{\mathrm{g}}$ and $E^{\mathrm{E}}_{\mathrm{g}}$),
as well as the width of the isolated metallic band ($W_{\mathrm{m}}$) obtained using the PBE 
functional with both FLEUR and \textsc{QuantumATK} codes. In the later code we employed both the LCAO and PW basis sets 
(the results using the PW-based \textsc{QuantumATK} are presented in parenthesis). 
While results from FLEUR and \textsc{QuantumATK} using the PBE functional 
show only minor absolute differences, these discrepancies can appear 
proportionally larger due to the small magnitudes of the values 
themselves. Similarly, within \textsc{QuantumATK}, the chosen basis set (LCAO 
or PW) also leads to small but observable variations. These findings 
underscore the sensitivity of numerical results to computational 
parameters, although among the most
recent electronic band structures there is a tendency to 
produce almost identical results when the same density functional is employed \cite{Science2016}. 
However, the differences arising from the choice of 
functional (e.g., PBE versus HSE06) are significantly larger and 
represent the core focus of this work, as they exert a far more 
substantial influence on the electronic properties relevant to device applications.
{This 
consistency in the absolute values across the differebt methods 
employed in our study suggests that the electronic properties derived with PBE are robust and largely 
unaffected by the choice of the \textit{ab initio} method. Generally, we observe that external 
band gaps are larger than internal gaps, though they typically remain around 1 eV, while the
width of the metallic band is often even greater than both
the internal and external energy gaps.

We can use the values of the energy gaps provided in Table \ref{table1} to deduce the cold metal character of each compound when the 
PBE functional is used. Since the external energy gap is greater than zero for all compounds under study, none is actually a n-type cold metal. 
Thus, when the internal energy gap is zero the compound is a p-type cold metal otherwise it is a pn-type cold metal. 
The majority of these compounds exhibit a pn-type cold metal character, characterized by an isolated metallic band at the Fermi level that is intersected by the Fermi 
energy (see Fig.\,\ref{fig1}). Typical band structures for NbS$_2$ and NbSi$_2$N$_4$, calculated
using both \textsc{QuantumATK} and FLEUR, are shown in Figs.\,\ref{fig4} and \ref{fig5}, respectively. 
As expected from the consistency observed in Table\,\ref{table1}, both \textit{ab initio} methods 
yield highly similar band structures under the PBE functional. A defining feature of these band 
structures is the single, isolated band at the Fermi level, distinctly separated from bands 
immediately below and above it. This aligns well with findings from Kuc et al., who studied the 
electronic properties of NbS$_2$ and other TMDCs, concluding that cold metallic behavior is present 
across its bulk, monolayer, bilayer, and quadrilayer forms \cite{Kuc2011}. Only a few compounds 
- NbTe$_2$ and NbSi$_2$P$_4$ - exhibit a p-type cold metal character. TaTe$_2$ is predicted to be a p-type cold metal when  \textsc{QuantumATK} is used while 
the FLEUR leads to a very small internal energy gap. Finally, we should note that TaSi$_2$P$_4$ exhibits a 
very small internal energy gap when PBE is employed in conjunction to both \textsc{QuantumATK} and FLEUR 
codes, which vanishes when $GW$ approximation is used as discussed 
below.

\begin{figure}
\begin{center}
\includegraphics[width=\columnwidth]{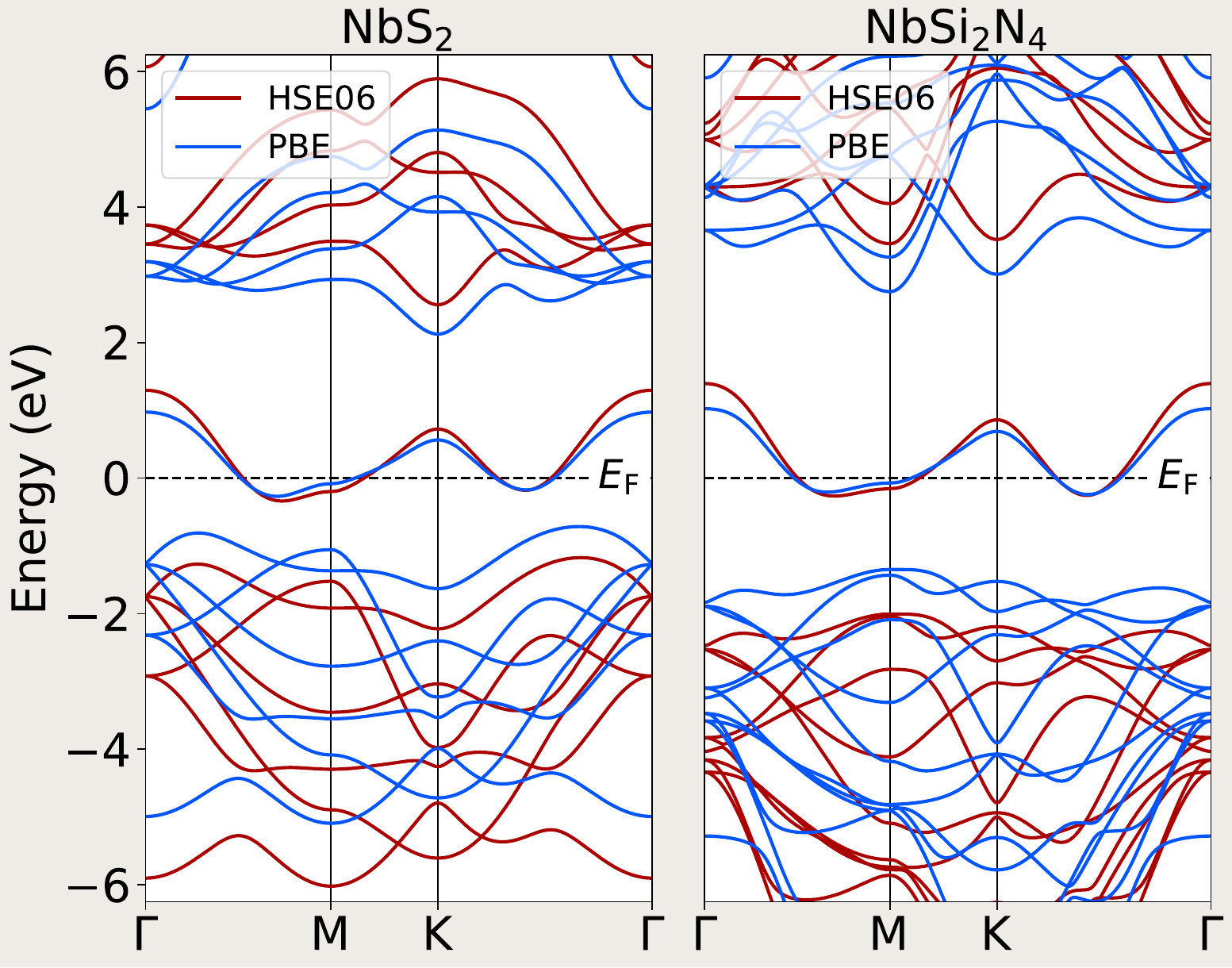}
\end{center}
\vspace*{-0.5cm}
\caption{Band structures of 1H NbS$_2$ and NbSi$_2$N$_4$ compounds along high-symmetry 
lines in the Brillouin zone, calculated using the LCAO-based
\textsc{QuantumATK} code with both the 
PBE (blue lines) and HSE06 (red lines) functionals.} 
\label{fig4}
\end{figure}

\subsection{Correlation effects: HSE06 hybrid functional and GW approximation}

Hybrid functionals like HSE06 are designed to provide a more accurate treatment 
of the exchange energy than standard GGA functionals, often resulting in an increased 
band gap in semiconductors. However, the effectiveness of these functionals varies based 
on material-specific properties due to the mixing coefficients used in combining 
exact Hartree-Fock and GGA exchange energies. In Ref. \onlinecite{Akanda2021} NbSi$_2$N$_4$ 
was studied using first-principles calculations with the HSE06 functional, revealing 
a pn-type cold metallic character. The authors showed that the isolated narrow band 
near the Fermi level is primarily a $d_{z^2}$-orbital band at the $\Gamma$ point, centered 
on the transition metal atoms (Nb). In contrast, the higher occupied valence bands are 
predominantly $p$-orbital bands from N atoms, while the lower unoccupied conduction 
bands are mainly derived from $d$ orbitals.

Table \ref{table1} presents our calculated results using the HSE06 functional. For all 
studied materials, HSE06 yields a moderate increase in both the internal and external 
energy gaps relative to the PBE functional, as expected, along with an increase in the 
width of the metallic band at the Fermi level. Notably, this effect is more pronounced 
in the p-type cold metals for which 
an internal energy gap emerges, classifying these materials as pn-type cold metals under 
HSE06. Figure\,\ref{fig4} compares the HSE06 and PBE band structures using the  LCAO-based version of the \textsc{QuantumATK} 
code, illustrating that while HSE06 does not alter the band shapes, it leads to a broader 
isolated band at the Fermi level and shifts the valence and conduction bands lower and higher 
in energy, respectively.

\begin{figure}
\begin{center}
\includegraphics[width=\columnwidth]{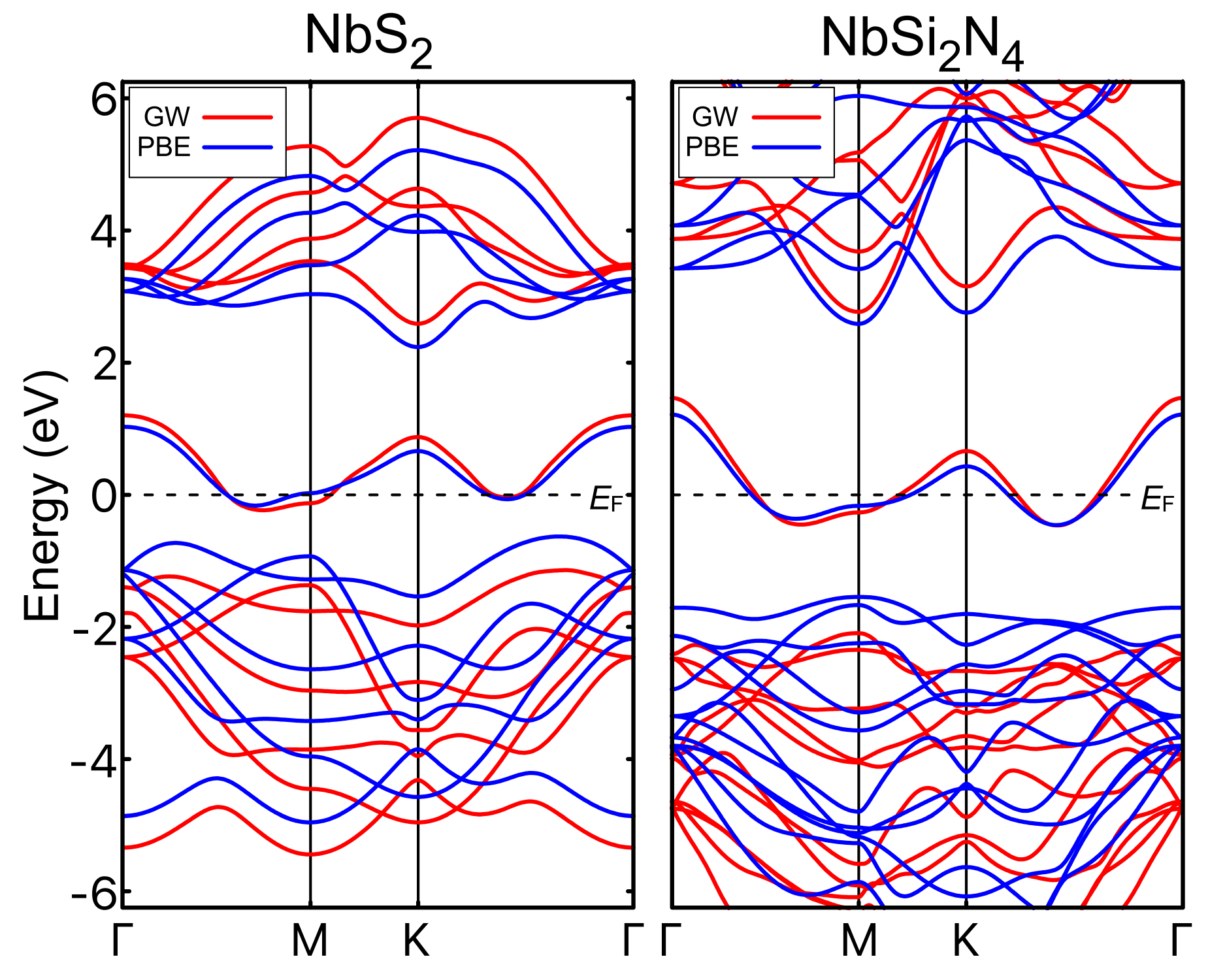}
\end{center}
\vspace*{-0.7 cm}
\caption{Band structures of 1H NbS$_2$ and NbSi$_2$N$_4$ compounds along high-symmetry lines 
in the Brillouin zone, calculated using the \textsc{FLEUR/SPEX} code with both the PBE (blue lines) 
and $GW$ approximation (red lines).} 
\label{fig5}
\end{figure}

Electron correlations are essential in determining the electronic structure of many 
materials. Although the correlation energy is relatively small—significantly less 
than the exchange energy and several orders of magnitude smaller than the Coulomb 
energy—it plays a crucial role in accurately describing electronic properties. Therefore,
methods like the $GW$ approximation are particularly valuable for studying the electronic 
structure of the materials in our research.

Kim et al. have reported the quasiparticle band structure of a monolayer of 1H-NbSe$_2$ 
using the Quantum Espresso method \cite{Kim2017}. Their approach involved a large simulation 
cell with a carefully converged interlayer distance to minimize interactions between periodic 
images of the monolayers. Their results showed that the $GW$ approximation produced a 
slightly broader band at the Fermi level compared to the PBE functional and led to significantly 
larger internal and external band gaps. Similar findings for NbS$_2$ monolayers were presented
in Ref. \onlinecite{Heil2018}, where $GW$ calculations closely matched ARPES experimental data, 
validating the accuracy of this approach.

\begin{figure}[t]
\begin{center}
\includegraphics[width=0.47\textwidth]{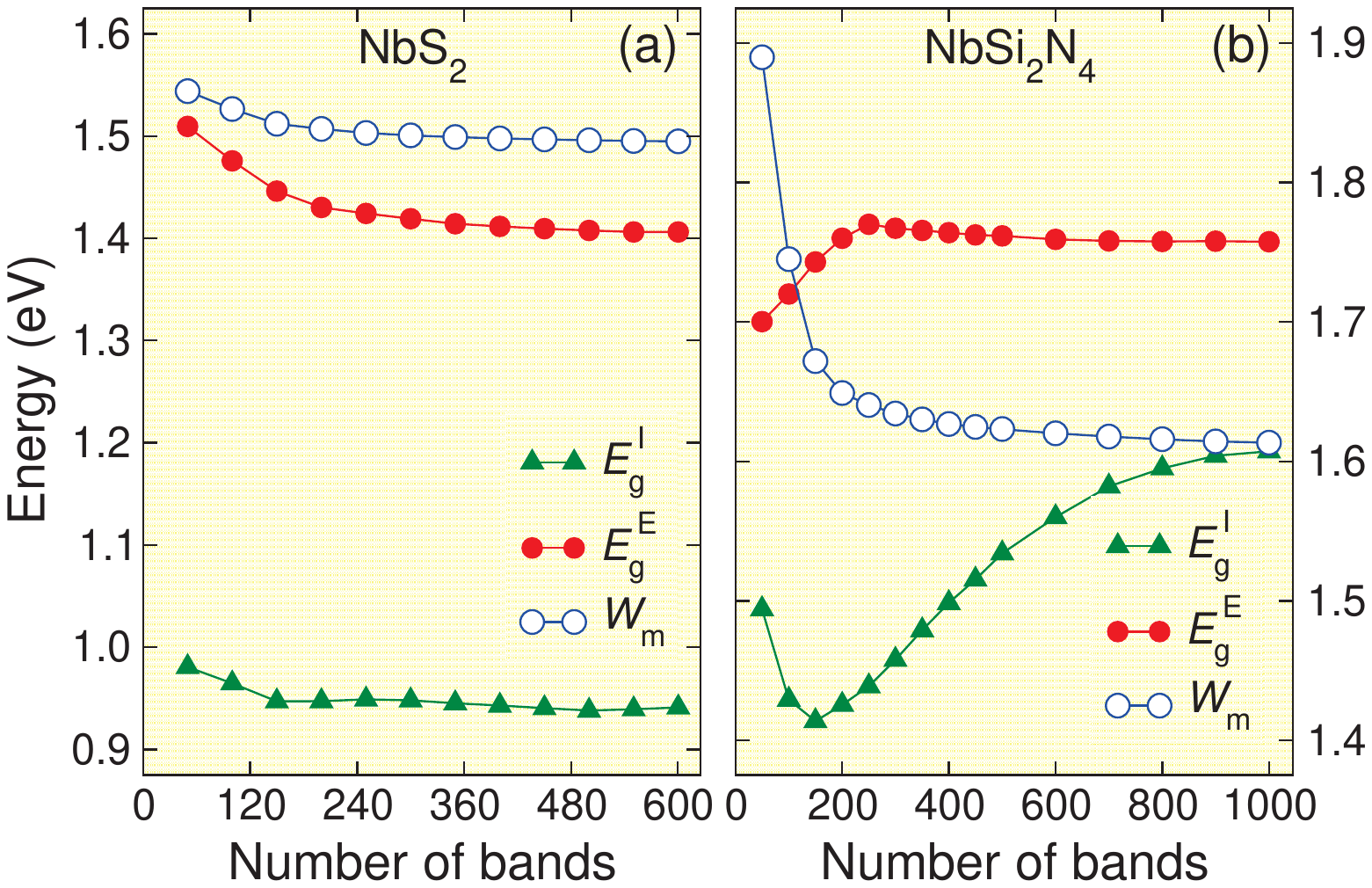}
\end{center}
\vspace*{-0.4cm}
\caption{Dependence of internal and external band gaps ($E^{\mathrm{I}}_{\mathrm{g}}$ and $E^{\mathrm{E}}_{\mathrm{g}}$), as well as the width $W_{\mathrm{m}}$ of the 
isolated metallic band, on the number of bands included in the $GW$ calculations (a) for
NbS$_2$ and  (b) for NbSi$_2$N$_4$.} \label{fig6}
\end{figure}

The importance of electronic correlations in NbS$_2$ monolayers was examined by 
Loon et al. \cite{Loon2018}, who investigated the combined effects of short- and 
long-range Coulomb interactions and electron–phonon coupling on the material's electronic 
properties. They found that the interplay of these three interactions generates electronic 
correlations that differ fundamentally from the effects of each interaction considered 
independently. The resulting fully interacting electronic spectral function closely 
resembles the non-interacting band structure but exhibits substantial broadening due 
to these correlations.

The first step in our $GW$ calculations involves ensuring convergence of the electronic 
band structure with respect to the number of bands included. In Fig. \ref{fig5}, we present 
convergence tests for two pn-type cold metals, NbS$_2$ and NbSi$_2$N$_4$, using a fixed 
\textbf{k}-point grid of $18 \times 18 \times 1$. For NbS$_2$, both the internal and external 
band gaps, as well as the width of the isolated metallic band, converge smoothly and rapidly 
as the number of bands increases, reaching stability within a few meV at approximately 120 bands. 
In contrast, NbSi$_2$N$_4$ presents a more challenging case, with the internal gap converging 
more slowly; achieving similar accuracy requires approximately 800 bands. Based on these 
findings, we used 600 bands for the $MX_2$ compounds and 1000 bands for the $MA_2Z_4$ compounds 
in our $GW$ calculations. Additionally, we performed convergence tests on the \textbf{k}-point 
sampling in the first Brillouin zone, confirming that our chosen grid provides energy gaps and 
bandwidths with accuracy within a few meV.

Table\,\ref{table1} presents the values of the external and internal band gaps, 
as well as the width of the isolated metallic band at the Fermi level, obtained 
using the $GW$ approximation. As expected, the $GW$ method yields larger band 
gaps and a broader metallic band compared to the PBE functional, exhibiting a trend 
similar to that observed with the HSE06 hybrid functional. However, there is no 
consistent pattern in the relative increases in gap values between $GW$ and HSE06: 
in some materials, HSE06 results in a greater increase over PBE, while in others, 
the $GW$ method shows a larger effect. The influence of $GW$ on the band structure 
of pn-type cold metals is illustrated in Fig.\,\ref{fig5}, where we compare the PBE 
and $GW$ band structures for NbS$_2$ and NbSi$_2$N$_4$. 
It is worth noting that 
non-self-consistent one-shot $GW$ calculations are known to not 
conserve the electronic charge, as the Green’s function and self-energy are not iteratively 
updated \cite{schindlmayr2001diagrammatic,aryasetiawan1998gw,onida2002electronic}. To address 
this issue, we used the DFT-PBE band structure as a reference for charge neutrality. Specifically,
we applied a uniform energy shift to the $GW$-calculated bands such that their intersection with the 
Fermi level closely matched the same  $\mathbf{k}$-points in the Brillouin zone as observed 
in the DFT-PBE band structure. This procedure ensured the 
charge neutrality in the $GW$ calculations and a consistent reference point for 
comparing the electronic properties of the materials under study.
The $GW$ approximation results in a broader metallic band at the Fermi level and 
simultaneously shifts the valence bands lower and the conduction bands higher in energy, 
leading to larger internal and external band gaps. Notably, the overall shape of the bands 
remains similar when comparing the PBE and $GW$ band structures. Overall, the impact of 
$GW$ on the PBE band structure resembles the effect observed with the HSE06 functional, 
as evident from the band structure comparisons in Figs.\,\ref{fig4} and \ref{fig5}.

A key distinction between HSE06 and $GW$ lies in their treatment of p-type cold metals.
As discussed, HSE06 yields a finite internal band gap for all  p-type materials 
in our study, indicating that these materials exhibit pn-type cold metallic behavior. 
In contrast, the $GW$ approximation for NbTe$_2$ and NbSi$_2$P$_4$ maintains the zero-gap 
character seen with PBE, thus preserving their p-type cold metallic nature. The behavior of the $GW$ approximation with respect to the PBE results, particularly the 
retention of the zero-gap character in p-type cold metals such as NbTe$_2$ and NbSi$_2$P$_4$, is 
expected given that we perform one-shot $GW$ calculations. In this approach, $GW$ acts as a perturbation to the PBE reference, without iterative updates to the Green's function or self-
energy. Consequently, the key characteristics of the PBE band structure are preserved. While fully 
self-consistent $GW$ calculations could potentially provide a more accurate description, they are 
computationally intensive and beyond the scope of the present study.
For TaSi$_2$P$_4$, 
$GW$ similarly closes the already minimal internal gap found in PBE, reinforcing its 
intrinsic zero-gap behavior. For TaTe$_2$ $GW$ keeps the pn-character observed when FLEUR in conjunction with PBE are employed. This discrepancy likely arises because HSE06 parameters 
are optimized primarily for semiconductors, limiting its general applicability across 
all material types. By contrast, $GW$ is a more universal approximation, treating correlation 
effects consistently across materials without altering their fundamental character.

\subsection{Implications for device applications}

The implications of our findings for electronic device applications are significant, particularly 
for the design of NDR tunnel diodes and steep-slope FETs. Among the 2D cold metals we investigated, 
some have already demonstrated potential as electrode materials in device simulations. In NDR tunnel 
diodes, three key electronic structure parameters—namely, the internal band gap ($E^{\mathrm{I}}_{\mathrm{g}}$), 
external band gap ($E^{\mathrm{E}}_{\mathrm{g}}$), and metallic bandwidth 
($W_{\mathrm{m}}$)—are crucial in determining the device’s $I$-$V$ characteristics and the PVCR value. 
To achieve a high PVCR in an N-type NDR effect, it is desirable that the metallic bandwidth 
$W_{\mathrm{m}}$ is smaller than both band gaps. If this condition is not met, PVCR may be 
significantly reduced. In $MX_2$  compounds, our GW approximation results reveal substantial 
increases in band gaps, especially for the internal gap. Despite this enhancement, the metallic 
bandwidth remains large. For instance, in NbS$_2$ and TaSe$_2$, $W_{\mathrm{m}}$ exceeds both 
$E^{\mathrm{I}}_{\mathrm{g}}$ and $E^{\mathrm{E}}_{\mathrm{g}}$ (i.e., $W_{\mathrm{m}} > 
E^{\mathrm{E}}_{\mathrm{g}} > E^{\mathrm{I}}_{\mathrm{g}}$), whereas, in NbSe$_2$ and TaSe$_2$, 
the external gap $E^{\mathrm{E}}_{\mathrm{g}}$ is greater than $W_{\mathrm{m}}$, followed by the 
internal gap ($E^{\mathrm{E}}_{\mathrm{g}}> W_{\mathrm{m}} > E^{\mathrm{I}}_{\mathrm{g}}$). 
The latter group is anticipated to produce an N-type NDR effect with moderate PVCR values when 
used as electrodes in tunnel diodes, while the former group may yield lower PVCR values. In particular,
NbSi$_2$N$_4$ emerges as a highly promising material for NDR tunnel diode applications, satisfying 
the ideal condition of $E^{\mathrm{E}}_{\mathrm{g}} > E^{\mathrm{I}}_{\mathrm{g}} > W_{\mathrm{m}}$. 
Prior studies (Ref. \onlinecite{Sasioglu2024}) have explored lateral NDR tunnel diodes based on 
NbSi$_2$N$_4$ electrodes, where DFT combined with non-equilibrium Green’s function (NEGF) simulations 
at the PBE level reported an impressively high PVCR of 10$^3$-10$^5$. Notably, at the PBE level, the 
internal band gap of NbSi$_2$N$_4$ is smaller than its bandwidth. Other materials in this group, 
such as NbGe$_2$N$_4$ and TaSi$_2$N$_4$, also hold promise, as indicated in Table\,\ref{table1} .

For FET applications, the critical parameters are the external band gap and the bandwidth 
$W_{\mathrm{m}}^{\mathrm{E}}$ above the Fermi level (often referred to as the valence band
maximum in p-type cold metals). A larger external band gap combined with a smaller bandwidth 
$W_{\mathrm{m}}^{\mathrm{E}}$ enhances the filtering of high-energy hot electrons in FETs, 
enabling subthreshold slopes below the thermionic limit of 60 mV/dec at room temperature. Recent 
studies have demonstrated sub-60 mV/dec SS values in FETs with cold metal electrodes like NbS$_2$, 
TaS$_2$, and NbTe$_2$ using the DFT+NEGF method at the PBE level \cite{Liu2020}. When correlation 
effects are incorporated in device simulations, similar subthreshold swings may be anticipated. 
However, while the increased external band gap due to correlations is beneficial, it may be 
counterbalanced by the concurrent increase in $W_{\mathrm{m}}^{\mathrm{E}}$, which could 
slightly reduce the efficiency of hot electron filtering. These insights underscore the promise 
and challenges of using 2D cold metals in advanced nanoelectronic devices, where a nuanced 
understanding of correlation effects is essential for optimizing device performance.

\section{Summary and Conclusions}

In this work, we investigated the influence of correlation effects on the electronic band structure
of 2D cold metals $MX_2$ and $MA_2Z_4$ ($M$=Nb, Ta; $X$=S, Se, Te; $A$=Si, Ge; $Z$=N, P) using the 
$GW$ approximation and HSE06 hybrid functional. These cold metals, characterized by their unique 
electronic features — namely, an isolated metallic band near the Fermi level with distinct internal 
and external band gaps — show substantial promise for use in advanced nanoelectronic devices, such as 
NDR tunnel diodes and FETs with steep subthreshold slopes. Our results indicate that both $GW$ and 
HSE06 enhance the internal ($E^{\mathrm{I}}_{\mathrm{g}}$) and external ($E^{\mathrm{E}}_{\mathrm{g}}$) 
band gaps and, to a lesser extent, the width of the metallic band ($W_{\mathrm{m}}$). However, the internal 
band gap $E^{\mathrm{I}}_{\mathrm{g}}$ displays the greatest sensitivity to correlation effects, 
underscoring its importance in determining the electronic characteristics of these materials. These
three band structure parameters are essential for optimizing the performance of cold metal-based devices.

For NDR tunnel diodes, our findings suggest that materials like NbSi$2$N$_4$, with an ideal band 
structure hierarchy ($E^{\mathrm{E}}_{\mathrm{g}} > E^{\mathrm{I}}_{\mathrm{g}} > W_{\mathrm{m}}$), 
are particularly suited for achieving high PVCR values.  Other compounds, such as NbGe$_2$N$_4$ and 
TaSi$_2$N$_4$, also show promising band structures conducive to NDR applications, with moderate 
PVCR values expected based on the band hierarchy. For FET applications, we highlight that a large 
external band gap combined with a narrow bandwidth above the Fermi level supports efficient hot-electron
filtering, which is necessary to achieve sub-60 mV/dec subthreshold swings at room temperature. Our study suggests that incorporating correlation effects in device simulations may reveal an improved capability for subthreshold performance,although the increased metallic bandwidth may present challenges for optimizing 
hot-electron filtering. Our findings provide a robust theoretical foundation for the use of 2D cold metals 
in nanoelectronic applications, where understanding and leveraging correlation effects is crucial. This 
work opens avenues for further experimental and theoretical research aimed at enhancing the design of 
cold metal-based devices, paving the way for next-generation electronics with superior performance 
metrics in both switching speed and energy efficiency.

\begin{acknowledgments}
This work was supported by several funding sources, including SFB CRC/TRR 227 and SFB 1238 (Project C01) 
of the Deutsche Forschungsgemeinschaft (DFG), the European Union (EFRE) through Grant No: ZS/2016/06/79307, 
and the Federal Ministry of Education and Research of Germany (BMBF) within the framework of the 
Palestinian-German Science Bridge (BMBF grant number DBP01436).

\end{acknowledgments}

\section*{Data Availability Statement}

Data available on request from the authors

\end{document}